\documentclass{ws-procs9x6}
\usepackage{feynarts,amsmath}

\newcommand{\gev}{\text{GeV}}
\newcommand{\tev}{\text{TeV}}

\newcommand{\pb}{\text{pb}}
\newcommand{\MSUSY}{ M_{\text{SUSY}}}

\begin{document}
\title{
Production of 
Neutral Higgs + Jet
at the LHC
in the SM and the MSSM\footnote{
\uppercase{T}alk presented by \uppercase{O.} \uppercase{B}rein
at {\it \uppercase{SUSY} 2003:
\uppercase{S}upersymmetry in the \uppercase{D}esert}\/, 
held at the \uppercase{U}niversity of \uppercase{A}rizona,
\uppercase{T}ucson, \uppercase{AZ}, \uppercase{J}une 5-10, 2003.
\uppercase{T}o appear in the \uppercase{P}roceedings.}}

\author{OLIVER BREIN}

\address{Institut f\"ur Theoretische Physik E, RWTH Aachen\\ 
Aachen, D-52056 Germany, E-mail: brein@physik.rwth-aachen.de}

\author{WOLFGANG HOLLIK}

\address{Max--Planck--Institut f\"ur Physik (Werner--Heisenberg--Institut)\\
M\"unchen, D-80805 Germany, E-mail: hollik@mppmu.mpg.de}


\maketitle

\abstracts{
The cross section prediction for the production of the lightest 
neutral Higgs boson in association 
with a high-$p_T$ hadronic jet is presented in the framework of 
the minimal supersymmetric standard model (MSSM)
and compared to the SM case.
Prospects for
the CERN Large Hadron Collider (LHC)
are discussed. 
}


\section{Higgs + Jet in the Standard Model}

The production of SM Higgs bosons in hadron collisions at the 
LHC will proceed mainly via gluon fusion ($gg\to H$).
The detection of a SM Higgs boson
with a mass below 140$\,\gev$
at the LHC is rather difficult
because the predominant decay
into a $b\bar b$-pair is swamped by the large QCD two-jet 
background \cite{ATLASTDR}.
Therefore, only through observation of the
rare decay into two photons,
is the inclusive single Higgs boson production 
considered the best search channel 
in this mass range at the LHC.

Alternatively,  
and in order to fully explore the Higgs-detection capabilities 
of the LHC detectors, 
one can investigate more exclusive channels 
like e.g.~Higgs production in association with a high-$p_T$ hadronic jet.
The main advantage of this channel is the richer kinematical structure 
of the events
which 
allows for refined cuts increasing the 
signal-to-background 
ratio.
Recently, a detailed 
simulation\cite{ADIKSS-leshouchesHWG} of the SM reaction
$pp\to H+\text{jet}+X, H\to\gamma\gamma$
using the basic cuts $p_{T,\text{jet}} \geq 30\,\gev$, 
$|\eta_{\text{jet}}|\leq 4.5$
showed that this process is a promising alternative or 
supplement to the inclusive Higgs-boson production for 
$m_H < 140\,\gev$.

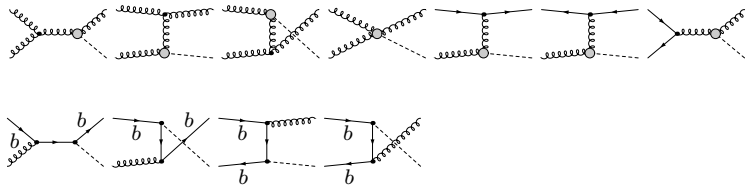
\begin{figure}[bt]
\begin{footnotesize}

\begin{picture}(378,40)(20,0)
\unitlength=1bp%
\begin{feynartspicture}(378,40)(7,1)
\FADiagram{}
\FAProp(0.,15.)(6.,10.)(0.,){/Cycles}{0}
\FAProp(0.,5.)(6.,10.)(0.,){/Cycles}{0}
\FAProp(20.,15.)(14.,10.)(0.,){/Cycles}{0}
\FAProp(20.,5.)(14.,10.)(0.,){/ScalarDash}{0}
\FAProp(6.,10.)(14.,10.)(0.,){/Cycles}{0}
\FAVert(6.,10.){0}
\FAVert(14.,10.){-1}

\FADiagram{}
\FAProp(0.,15.)(10.,14.)(0.,){/Cycles}{0}
\FAProp(0.,5.)(10.,6.)(0.,){/Cycles}{0}
\FAProp(20.,15.)(10.,14.)(0.,){/Cycles}{0}
\FAProp(20.,5.)(10.,6.)(0.,){/ScalarDash}{0}
\FAProp(10.,14.)(10.,6.)(0.,){/Cycles}{0}
\FAVert(10.,14.){0}
\FAVert(10.,6.){-1}

\FADiagram{}
\FAProp(0.,15.)(10.,14.)(0.,){/Cycles}{0}
\FAProp(0.,5.)(10.,6.)(0.,){/Cycles}{0}
\FAProp(20.,15.)(10.,6.)(0.,){/Cycles}{0}
\FAProp(20.,5.)(10.,14.)(0.,){/ScalarDash}{0}
\FAProp(10.,14.)(10.,6.)(0.,){/Cycles}{0}
\FAVert(10.,6.){0}
\FAVert(10.,14.){-1}

\FADiagram{}
\FAProp(0.,15.)(10.,10.)(0.,){/Cycles}{0}
\FAProp(0.,5.)(10.,10.)(0.,){/Cycles}{0}
\FAProp(20.,15.)(10.,10.)(0.,){/Cycles}{0}
\FAProp(20.,5.)(10.,10.)(0.,){/ScalarDash}{0}
\FAVert(10.,10.){-1}

\FADiagram{}
\FAProp(0.,15.)(10.,14.)(0.,){/Straight}{1}
\FAProp(0.,5.)(10.,6.)(0.,){/Cycles}{0}
\FAProp(20.,15.)(10.,14.)(0.,){/Straight}{-1}
\FAProp(20.,5.)(10.,6.)(0.,){/ScalarDash}{0}
\FAProp(10.,14.)(10.,6.)(0.,){/Cycles}{0}
\FAVert(10.,14.){0}
\FAVert(10.,6.){-1}

\FADiagram{}
\FAProp(0.,15.)(10.,14.)(0.,){/Straight}{-1} 
\FAProp(0.,5.)(10.,6.)(0.,){/Cycles}{0}
\FAProp(20.,15.)(10.,14.)(0.,){/Straight}{1}
\FAProp(20.,5.)(10.,6.)(0.,){/ScalarDash}{0}
\FAProp(10.,14.)(10.,6.)(0.,){/Cycles}{0}
\FAVert(10.,14.){0}
\FAVert(10.,6.){-1}

\FADiagram{}
\FAProp(0.,15.)(6.,10.)(0.,){/Straight}{1}
\FAProp(0.,5.)(6.,10.)(0.,){/Straight}{-1}
\FAProp(20.,15.)(14.,10.)(0.,){/Cycles}{0}
\FAProp(20.,5.)(14.,10.)(0.,){/ScalarDash}{0}
\FAProp(6.,10.)(14.,10.)(0.,){/Cycles}{0}
\FAVert(6.,10.){0}
\FAVert(14.,10.){-1}
\end{feynartspicture}
\end{picture}

\begin{picture}(216,40)
\begin{feynartspicture}(216,40)(4,1)
\FADiagram{}
\FAProp(0.,15.)(6.,10.)(0.,){/Straight}{1}
\FALabel(2.48771,11.7893)[tr]{$b$}
\FAProp(0.,5.)(6.,10.)(0.,){/Cycles}{0}
\FAProp(20.,15.)(14.,10.)(0.,){/Straight}{-1}
\FALabel(16.4877,13.2107)[br]{$b$}
\FAProp(20.,5.)(14.,10.)(0.,){/ScalarDash}{0}
\FAProp(6.,10.)(14.,10.)(0.,){/Straight}{1}
\FAVert(6.,10.){0}
\FAVert(14.,10.){0}

\FADiagram{}
\FAProp(0.,15.)(10.,14.)(0.,){/Straight}{1}
\FALabel(4.84577,13.4377)[t]{$b$}
\FAProp(0.,5.)(10.,6.)(0.,){/Cycles}{0}
\FAProp(20.,15.)(10.,6.)(0.,){/Straight}{-1}
\FALabel(16.8128,13.2058)[br]{$b$}
\FAProp(20.,5.)(10.,14.)(0.,){/ScalarDash}{0}
\FAProp(10.,14.)(10.,6.)(0.,){/Straight}{1}
\FAVert(10.,14.){0}
\FAVert(10.,6.){0}

\FADiagram{}
\FAProp(0.,15.)(10.,14.)(0.,){/Straight}{1}
\FALabel(4.84577,13.4377)[t]{$b$}
\FAProp(0.,5.)(10.,6.)(0.,){/Straight}{-1}
\FALabel(5.15423,4.43769)[t]{$b$}
\FAProp(20.,15.)(10.,14.)(0.,){/Cycles}{0}
\FAProp(20.,5.)(10.,6.)(0.,){/ScalarDash}{0}
\FAProp(10.,14.)(10.,6.)(0.,){/Straight}{1}
\FAVert(10.,14.){0}
\FAVert(10.,6.){0}

\FADiagram{}
\FAProp(0.,15.)(10.,14.)(0.,){/Straight}{1}
\FALabel(4.84577,13.4377)[t]{$b$}
\FAProp(0.,5.)(10.,6.)(0.,){/Straight}{-1}
\FALabel(5.15423,4.43769)[t]{$b$}
\FAProp(20.,15.)(10.,6.)(0.,){/Cycles}{0}
\FAProp(20.,5.)(10.,14.)(0.,){/ScalarDash}{0}
\FAProp(10.,14.)(10.,6.)(0.,){/Straight}{1}
\FAVert(10.,14.){0}
\FAVert(10.,6.){0}
\end{feynartspicture}
\end{picture}

\end{footnotesize}

\caption{\label{SM_parton_processes}
Partonic processes contributing to 
$pp \to H + \text{jet} + X$ in the SM.
Hatched circles represent
loops of heavy quarks. The depicted tree-level $b$-quark processes
are much more important in the MSSM case.}
\end{figure}

The partonic processes,
calculated at present to leading order\cite{hjet-sm},
contributing to the hadronic reaction 
$pp \to H + \text{jet} + X$ (see Fig.~\ref{SM_parton_processes}) 
are gluon fusion ($gg \to g H$, 50--70 \% of total rate), 
quark--gluon scattering 
($q(\bar q)g \to q(\bar q) H$, 30--50 \% of total rate)
and 
quark--antiquark annihilation ($q\bar q \to g H$, rate small).
The hadronic 
cross section is dominated by loop-induced processes, involving
effective $ggH$-- and $ggHZ$--couplings. If the $b$-quark is treated as a 
parton present in the proton, there are additional tree-level processes
for quark-gluon scattering and quark-antiquark annihilation
to consider. Yet, in the SM their contribution to the hadronic cross
section is small.

\section{Higgs + Jet in the MSSM}

Motivated by the promising SM simulation\cite{ADIKSS-leshouchesHWG}
we investigated\cite{hjet-paper} the MSSM process $pp\to h^0+\text{jet}+X$,
involving the lightest MSSM Higgs boson.
Especially, as the process is essentially 
loop-induced, there are potentially large effects from virtual
superpartners to be expected.

In the MSSM, the contributing partonic processes are 
basically the same as in the SM,
{\it i.e.}~$gg\to gh^0, 
q(\bar q)\to q(\bar q)h^0, q\bar q\to g h^0$, 
but there are differences resulting from various sources: 
(i) different Yukawa couplings in the MSSM
involving the mixing angles $\alpha$ and $\beta$ of the MSSM Higgs sector.
Especially, 
the $b$-quark processes are enhanced for small 
values of the CP-odd Higgs boson mass $m_A$ ($< 150\,\gev$)
and large $\tan\beta$ in the MSSM.
(ii) Additional superpartner contributions
to the amplitudes, squark-loop insertions
for all SM-like topologies displayed in Fig.~\ref{SM_parton_processes}
and new Feynman graph topologies containing at least
one gluino line (Fig.~2). 
The calculation of the 
partonic processes was done with the help of FeynArts/FormCalc\cite{FAFC}.

\begin{figure}[bt]
\begin{footnotesize}
\begin{picture}(378,40)(20,0)
\unitlength=1bp%
\begin{feynartspicture}(378,40)(6,1)
\FADiagram{}
\FAProp(0.,15.)(6.,10.)(0.,){/Straight}{1}
\FAProp(0.,5.)(6.,10.)(0.,){/Cycles}{0}
\FAProp(20.,15.)(14.,10.)(0.,){/Straight}{-1}
\FAProp(20.,5.)(14.,10.)(0.,){/ScalarDash}{0}
\FAProp(6.,10.)(14.,10.)(0.,){/Straight}{1}
\FAVert(6.,10.){0}
\FAVert(14.,10.){-1}

\FADiagram{}
\FAProp(0.,15.)(10.,14.)(0.,){/Straight}{1}
\FAProp(0.,5.)(10.,6.)(0.,){/Cycles}{0}
\FAProp(20.,15.)(10.,6.)(0.,){/Straight}{-1}
\FAProp(20.,5.)(10.,14.)(0.,){/ScalarDash}{0}
\FAProp(10.,14.)(10.,6.)(0.,){/Straight}{1}
\FAVert(10.,6.){0}
\FAVert(10.,14.){-1}

\FADiagram{}
\FAProp(0.,15.)(10.,10.)(0.,){/Straight}{1}
\FAProp(0.,5.)(10.,10.)(0.,){/Cycles}{0}
\FAProp(20.,15.)(10.,10.)(0.,){/Straight}{-1}
\FAProp(20.,5.)(10.,10.)(0.,){/ScalarDash}{0}
\FAVert(10.,10.){-1}

\FADiagram{}
\FAProp(0.,15.)(10.,14.)(0.,){/Straight}{1}
\FAProp(0.,5.)(10.,6.)(0.,){/Straight}{-1}
\FAProp(20.,15.)(10.,14.)(0.,){/Cycles}{0}
\FAProp(20.,5.)(10.,6.)(0.,){/ScalarDash}{0}
\FAProp(10.,14.)(10.,6.)(0.,){/Straight}{1}
\FAVert(10.,14.){0}
\FAVert(10.,6.){-1}

\FADiagram{}
\FAProp(0.,15.)(10.,14.)(0.,){/Straight}{1}
\FAProp(0.,5.)(10.,6.)(0.,){/Straight}{-1}
\FAProp(20.,15.)(10.,6.)(0.,){/Cycles}{0}
\FAProp(20.,5.)(10.,14.)(0.,){/ScalarDash}{0}
\FAProp(10.,14.)(10.,6.)(0.,){/Straight}{1}
\FAVert(10.,14.){0}
\FAVert(10.,6.){-1}

\FADiagram{}
\FAProp(0.,15.)(10.,10.)(0.,){/Straight}{1}
\FAProp(0.,5.)(10.,10.)(0.,){/Straight}{-1}
\FAProp(20.,15.)(10.,10.)(0.,){/Cycles}{0}
\FAProp(20.,5.)(10.,10.)(0.,){/ScalarDash}{0}
\FAVert(10.,10.){-1}

\end{feynartspicture}
\end{picture}
\end{footnotesize}

\caption{\label{MSSM-xtra-tops}
Additional topologies for the loop-induced 
process $q g \to q h^0$ and $q\bar q\to g h^0$ in the MSSM. The hatched circles
represent loops containing at least one gluino line. 
}
\end{figure}
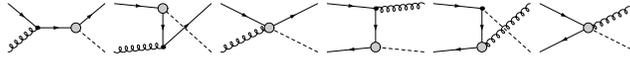

\section{Results}

The partonic cross sections $\hat\sigma_{n m}(\hat s)$ determine the 
hadronic cross section for $pp\to h^0+\text{jet}+X$ via the convolution
\begin{align*}
\sigma & =
\int_{\tau_0}^{1} d\tau 
\bigg(
\frac{ d{L}^{p p}_{g g} }{ d\tau }
\;\hat\sigma_{g h^0}(\tau S)
+\sum_{q} \frac{ d{L}^{p p}_{q g} }{ d\tau }
\;\hat\sigma_{q h^0}(\tau S)
+\sum_{q} \frac{ d{L}^{p p}_{q \bar q} }{ d\tau }
\;\hat\sigma_{g h^0}(\tau S)
\bigg)
\; ,
\end{align*}
with parton luminosities ${d{L}^{AB}_{nm}}/{d\tau}$,
and $\sqrt{S}=14\,\tev$ for the LHC. 
The cuts $p_{T,\text{jet}} \geq 30\,\gev$ and $|\eta_{\text{jet}}| \leq 4.5$
specify $\tau_0$ and the angular-integration limits;
they are chosen as in the SM analysis\cite{ADIKSS-leshouchesHWG}.

\begin{figure}[t]
{\setlength{\unitlength}{1cm}
\begin{picture}(4.5,6.)
\put(-1,0){\includegraphics{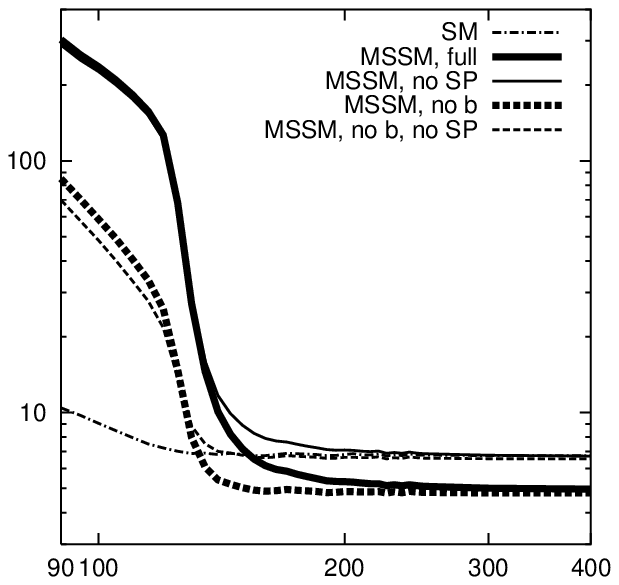}}
\put(1.4,2.15){{\small (a)}}
\put(3,1.35){$\scriptstyle m_A [\gev]$}
\put(0.45,4){\footnotesize $\sigma [\pb]$}
\put(2.9,4.7){\footnotesize $m_h^{\text{max}}$ scenario}
\put(2.9,4.3){\footnotesize $\MSUSY=400\,\gev$}
\put(2.9,3.9){\footnotesize $\tan\beta=30$}
\end{picture}
\begin{picture}(4.5,6.)
\put(0,0){\includegraphics{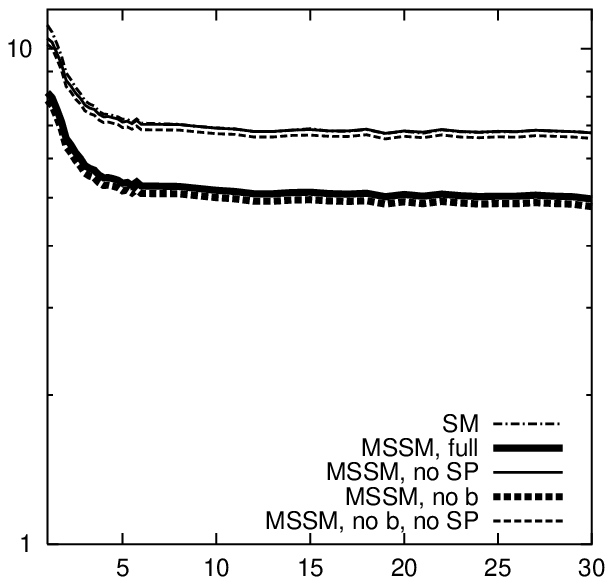}}
\put(2.3,2.15){{\small (b)}}
\put(4.,1.35){$\scriptstyle \tan\beta$}
\put(1.35,4){\footnotesize $\sigma [\pb]$}
\put(3.9,4.0){\footnotesize $m_h^{\text{max}}$ scenario}
\put(3.9,3.6){\footnotesize $\MSUSY=400\,\gev$}
\put(3.9,3.2){\footnotesize $m_A = 400\,\gev$}
\end{picture}\\
\begin{picture}(3.5,4.3)
\put(-.9,-.8){\includegraphics{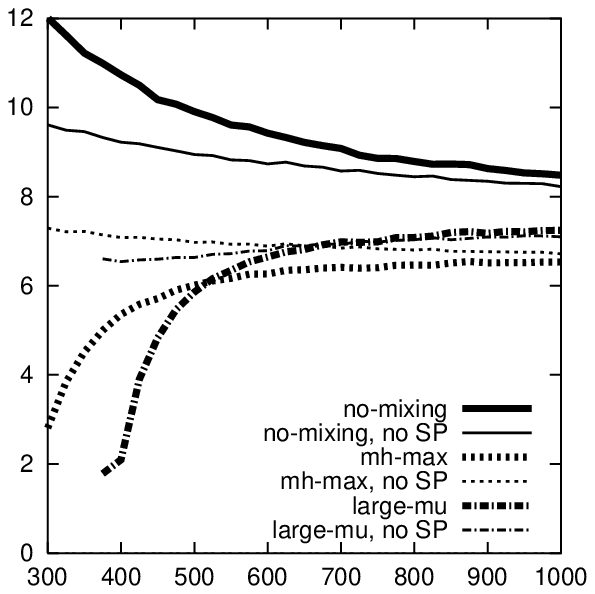}}
\put(1.35,1.3){{\small (c)}}
\put(2.5,.3){$\scriptstyle M_{\text{SUSY}} [\gev]$}
\put(0.42,3.53){\footnotesize $\sigma [\pb]$}
\put(3.3,5.){\footnotesize $m_A = 200\,\gev$}
\put(3.3,4.6){\footnotesize $\tan\beta=6$}
\end{picture}
\begin{picture}(5,4.3)
\put(0,-2.45){\includegraphics{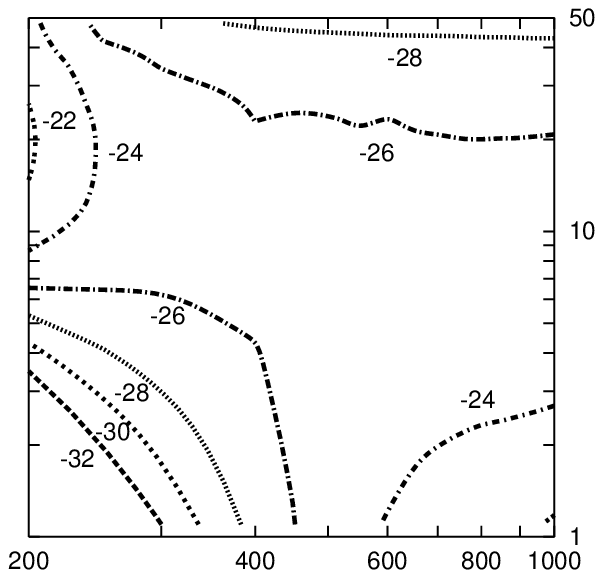}}
\put(3.45,1.25){{\small (d)}}
\put(5,.3){$\scriptstyle m_A [\gev]$}
\put(7.8,2.7){$\scriptstyle \tan\beta$}
\put(4.15,3.6){\footnotesize $(\sigma_{\text{MSSM}}-\sigma_{\text{SM}})/\sigma_{\text{SM}}\; [\%]$}
\put(5.05,3.2){\footnotesize $m_h^{\text{max}}$ scenario}
\put(5.05,2.8){\footnotesize $\MSUSY=400\,\gev$}
\end{picture}
}
\caption{\label{results} 
Hadronic cross sections in picobarn
(a) as a function of $m_A$,  
(b) as a function of $\tan\beta$, 
both in the $m_h^{\text{max}}$ scenario,
(c) as function of $\MSUSY$ for the three benchmark scenarios.
(d) Relative difference of MSSM and SM prediction in the $m_h^{\text{max}}$
scenario.}
\end{figure}

Figs.~\ref{results}(a),(b) show results for the MSSM $m_h^{\text{max}}$ 
benchmark scenario\cite{improvedbm}
with a common squark-mass scale $\MSUSY=400\,\gev$.
For small values of the CP-odd Higgs boson mass $m_A$ ($< 150\,\gev$) and especially for large $\tan\beta$, 
the $b$-quark Yukawa coupling 
is strongly enhanced compared to the top-Higgs coupling.
Thus, in this parameter range, 
the $b$-quark processes dominate the hadronic cross section,
and also the loop-induced processes are dominated by the $b$-quark loops.
At large $m_A$ ($> 200\,\gev$), the coupling of the $b$--Higgs
coupling is much smaller than the top--Higgs coupling and therefore 
the loop-induced processes
dominate.
For $m_A = 400\,\gev$, and almost independently of $\tan\beta$,
the full result is reduced by 24\%
compared to the result with quark loops only. 
Fig.~\ref{results}(c) shows the $\MSUSY$--dependence of
the hadronic cross section with and without 
superpartner contributions for the three benchmark
scenarios\cite{improvedbm}.
(The large-$\mu$ scenario had to be modified\cite{hjet-paper} in order to obey
the exclusion limit for the Higgs mass set by LEP data.)
For moderate, but allowed, squark masses the contribution of 
superpartners is significant.
In Fig.~\ref{results}(d) we display the relative difference
between the MSSM $m_h^{\text{max}}$ scenario and the 
SM prediction of the hadronic cross section 
plotted versus $m_A$ and $\tan\beta$.
The MSSM prediction 
is more than 20\% below the SM result 
(using the MSSM value of $m_{h^0}$ for the SM Higgs mass) 
in the 
whole area of the $m_A$--$\tan\beta$ plane displayed. 

In summary, Higgs + jet production 
is promising and deserves
a closer 
look in LHC physics simulations. 
To that end we provide the 
FORTRAN code 
for SM and MSSM predictions
for general use.

\end{document}